\documentstyle{article}

\begin{document}

\begin{center}

{\large 
 On the mechanism of low-mass compact object formation} \\

\vskip 0.1cm
{\bf S.B. Popov}

\vskip 0.1cm
University of Padova, Italy

\vskip 0.1cm
Sternberg Astronomical Institute, Russia 

\vskip 0.1cm
\vskip 0.1cm

{\bf Abstract}

\vskip 0.9cm

\end{center}  

{\it  We suggest that low-mass compact objects (hadron stars, quark stars)
with $M<1\,M_{\odot}$ can appear only due to fragmentation of  rapidly
rotating proto-neutron stars. 
Such low-mass stars receive  large kicks due to an explosion of a lighter
companion in a pair of fragments, or due to dynamical ejection of one of
the lighter components in the case when three bodies are formed.  

As far as low-mass compact objects 
are expected to be slowly cooling in all popular models of thermal evolution
possible candidates are expected to be found among hot
high velocity sources.}

\vskip 0.2cm

\vfill
{\it This short note is posted on astro-ph only,
and it will not be submitted to any paper journal.}

\newpage

In many models of thermal evolution 
of compact objects (neutron stars -- NS,
hybrid stars -- HyS, strange stars, -- SS) low-massive sources with 
$0.8\, M_{\odot} <M<1\, M_{\odot}$ 
\footnote{Here and below speaking about compact objects we mean the
gravitational mass.} 
remain hot for a relatively long time (about few million years) 
\footnote{Objects with even lower mass, $\sim 0.5\, M_{\odot}$,
 also are relatively hot (Blaschke et al. 2004).}.
During all that time they remain hotter 
than more massive stars  (Blaschke et al. 2004 and references therein). 
In that sense they are promising candidates to be
observed as {\it coolers}, and their detection is of great interest for
physics of dense matter (Carriere et al. 2003). 
However, in most of models of NS formation
(Woosley et al. 2002; Fryer, Kalogera 2001, Timmes et al. 1996) no objects
with $M<1$~--~1.2~$M_{\odot}$ are formed. It is like that 
just because masses of
stellar cores are always heavier than 1.2~$M_{\odot}$ even for the solar
metallicity (and heavier for lower metallicity, see Woosley et al. 2002
figs. 14 and 15).

In our opinion
the only way to form a low-mass object from a (relatively) high-mass core   
is fragmentation (see however a discussion in Xu 2004, where the author
discuss formation of low-mass SSs from white dwarfs via accretion induced
collapse).

 Fragmentation of a rapidly rotating protoNS due to 
dynamical instabilities as part of a two-stage supernova
(SN) explosion mechanism was suggested by Imshennik (1992)
(see also Imshennik, Nadezhin 1992).\footnote{Note that the English-language
versions of these papers are available via NASA ADS.} 
This mechanism can explain several
particular features of SN explosions (for example delay in
neutrino signal from SN1987A, see a recent discussion of this topic in
Imshennik, Ryazhskaja 2004). In this scenario a compact object inevitably
obtains a high kick velocity.  
Recently the mechanism (in application to kick)
was studied in some details by Colpi and Wasserman (2002).    
Because of high temperature the exploding (lighter) companion
can be significantly more massive than the minimum mass for cold stars 
(i.e. $\sim 0.1\, M_{\odot}$) up to $\sim 0.7 \,M_{\odot}$ (see Colpi,
Wasserman 2002). 
In that case from the initial object of, say,
1.2~--~1.3 solar masses due to fragmentation and explosion of the lighter part,
we can finally have a high-velocity NS with the
mass about 0.8-1 solar masses or
 lower if the mass of the initial object was smaller.

 Even lower masses can appear if due to fragmentation three bodies are
formed. In that case the lightest or an
intermediate mass fragment can be dynamically ejected
from the system (again with significant velocity about thousands km~s$^{-1}$). 
Such ejected compact objects can have masses about 0.2--0.5 solar masses.
In the remaining pair the lighter one can start to accrete onto the second
companion because of the orbit shrinking due to gravitational waves emission, 
and after reaching the minimum mass ($\sim 0.1\, M_{\odot}$)
it explodes. So, the remaining compact object would also have relatively
low mass ($\sim 1\, M_{\odot}$) and high velocity. 

 Objects formed after fragmentation have particular predictable properties:
high spatial velocity, high surface temperature, velocity vector nearly
perpendicular to spin axis (as far as kick is always obtained in the orbital
plane which coincides with the equator of the initial protoNS). 

 Due to high kicks low-mass compact objects are not expected to appear in
binaries (at least they should be rarely found in binary systems).
 To find a low-mass compact object one has to search for a
hot young high velocity NS. 

The best place to look for young hot compact objects are supernova remnants
(SNRs). 
Several so-called central compact objects (CCOs) are known (see  reviews in
Becker, Aschenbach 2002 and in Kaplan et al. 2004).
For example Cas A, Puppis A and 1E 1207.4-5209. The last one was suggested
to be a low-mass quark star candidate by Xu (2004). Cas A is known to be a
mysterious source with very small estimated emitting area and relatively
high temperature (0.7 keV), see Chakrabarty et al. (2001), Pavlov et al.
(2000). Puppis A is know to be a very hot source for its age (0.4 keV at 3.7
kyr, see Kaplan et al. 2004). 
However in the case of low-mass NSs
 the extremely high kick velocity can spoil this approach of searching.
If kick velocity is up to 10,000 km~s$^{-1}$ (as it is estimated in Colpi,
Wasserman 2002 for their case 2, i.e. for lighter component with the mass
$\sim 0.3\, M_{\odot}$) then a NS travels up to 1 kpc in just 100,000 years!
In that case they quickly leave the Galaxy, and of course there is little
hope to find them inside SNRs.

There is a possibility, that a rapidly rotating protoNS can just
loose part of its mass in the form of an  outflow in the equatorial plane.
In that case two spiral arms appear, no
second (or third) component is formed, and kick can be relatively small,  
but the fraction of lost mass is very small: about 4\% (Houser et al. 1994),
and so the final mass cannot be much lower than the initial one. 

 It is reasonable to expect that mass and kick velocity are
anti-correlated, as far as a higher mass of the remaining object
corresponds to the lighter exploded
component, which means to a wider orbit, and  to lower orbital velocity
of the remaining more massive component. Also higher kicks lead to smaller
fall-back (Colpi, Wasserman 2002).

 To reach fragmentation conditions (the dynamical  instability)
it is necessary that progenitor core
is rapidly rotating. Rotation of isolated progenitors and its influence on
properties of newborn NSs was studied in several papers (see, for example,
Heger et al. 2003 and references there in).  
To obtain a rapidly rotating compact object it is necessary to avoid
spin-down influence of the magnetic field, so probably compact objects born
after fragmentation should be low magnetized. It means that low-mass neutron
stars are not expected to be normal radio pulsars. Because of the same
reason they are not expected to show any kind of
magnetar activity.

It is interesting to investigate the possibility of a fossil disk formation
around a compact object which experienced fragmentation.
Very low masses of discs (about 0.1~$M_{\odot}$ and lower, Menou et al. 2001) 
are necessary in  models which explain SGRs and AXPs properties, and such
small amounts of mass are definitely available after fragmentation.  

 Evolution of progenitors in close binaries was  studied by several
groups (see Langer et al. 2003, Podsiadlowski et al. 2003 and references
there in). There are arguments both {\it pro et contra}
formation of rapidly rotating protoNSs in close binaries.
Iben and Tutukov (1996) discussed the possibility that fast rotation of a
compact object can be reached  only in close
binaries due to tidal corotation or due to accretion.
Langer et al. (2003) showed, that the primary (more massive) star in the
binary can hardly produce a rapidly rotating remnant as far as such stars
loose angular momentum due to mass loss. On the other hand these authors
suggested, that the secondary (which accretes matter and so angular
momentum) can be a progenitor of a short period compact object.
So, progenitors of low-mass NSs can be secondary components of close
binaries. It gives an opportunity to estimate the formation rate of such
objects. 

 There is a possibility that a low-mass compact object can appear to be not 
a normal (hadron) NS, but a quark star (a hybrid one or a bare strange).
The phase transition can appear long time after a NS formation or during the
very first episodes of NS's life 
(see for example Olinto 1987). 
One can think about a rather exotic possibility of phase transition of
newborn fragments into quark phase due to compression (increase of the
central density) observed in numerical
experiments by Mathews and Wilson (2000). Initial results of these authors
(see Mathews, Wilson 1997 and references to earlier papers there in)
were found to be incorrect (Flanagan 1999), 
but re-calculated models were not under serious
criticism and they still showed a moderate level compression  during NS-NS
coalescence
(see also Gourgoulhon et al. 2001, Oeshlin et al. 2002), 
but the degree of compression is close to numerical 
accuracy of calculations. 
In addition to these uncertainties in the case of protoNSs fragmentation
high temperatures and fast rotation
of new-formed fragments can exclude high densities
which are necessary for the phase transition.
Anyway, to form a low-mass quark star it is necessary at first 
to form a low-mass
NSs and here, in our opinion, fragmentation is the only mechanism,
but the role of compression is unclear (and probably negligible). 
One alternative to compression 
can be a delayed phase transition in a
low-mass NS  due to formation of a critical-size drop of strange matter
 (Berezhiani et al. 2003), but the "waiting time" 
(mean-life time of a metastable configuration) for a low-mass
object is estimated to be very long (Bombaci et al. 2004).
Another alternative is a strangelet passage through a
star (see however a discussion in Balberg 2004). 
Such transition to quark matter can also produce
additional energy release.
These objects are particularly interesting as far as low-massive HySs and
SSs
in contrast with low-mass hadron NSs can have relatively small radii, and
this feature can help in fitting observational data  
(see for example Xu 2004 for a discussion). 
\footnote{Phase transitions due to compression can be discussed also in the
standard scenario of binary NS coalescence.}

We have to note, that the mechanism of SN explosion
suggested by Imshennik (1992)
has its internal problems. If the fragmentation in the process of NS
formation never happens in Nature, then, in our opinion, it is very
improbable, that low-mass compact objects can exist.
Discovery of a high velocity low-mass NS, HyS 
or SS will be a strong argument in
favour of the Imshennik mechanism.   

\bigskip

To conclude: fragmentation of a protoNS can be a unique mechanism of 
the formation of low-mass compact objects, which are expected to have
several peculiar characteristics that can help to distinguish them among 
possible candidates.

\vskip 1.1cm

It is a pleasure to thank V.S. Imshennik and M.E. Prokhorov for discussions
and S. Rosswog for comments on the compression effect.




\newpage

\begin{thebibliography}{}


\bibitem{}Becker, W., Aschenbach, B., 2002,
in:  ``WE-Heraeus seminar on neutron stars,  
pulsars, and supernova remnants'', MPE Report 278, Eds. W.
            Becker, H. Lesch, J. Tr\"umper, Garching bei M\"unchen:
Max-Plank-Institut f\"ur extraterrestrische Physik, p.64
(astro-ph/0208466)

\bibitem{}Balberg, S., 2004,  Phys. Rev. Lett. 92,  119001
(astro-ph/0403503)

\bibitem{}Berezhiani, Z., Bombaci, I., Drago, A. et al., 2003,
ApJ 586, 1250

\bibitem{}Blaschke, D., Grigorian, H., Voskresenky, D.N., 2004, 
astro-ph/0403170

\bibitem{}Bombaci, I., Parenti, I., Vidana, I., 2004, astro-ph/0402404

\bibitem{}Carriere, J., Horowitz, C.J., Piekarewicz, J.
2003, ApJ 593, 463

\bibitem{}Chakrabarty, D., Pivovaroff, M.J. Hernquist, L.E., et al., 2001, 
ApJ 548, 800

\bibitem{}Colpi, M., Wasserman, I., 2002,
ApJ 581, 1271 (astro-ph/0207327) 

\bibitem{}Flanagan, E., 1999, Phys. Rev. Lett. 82, 1354

\bibitem{}Fryer, C.L., Kalogera, V., 2001, ApJ 554, 548

\bibitem{} Gourgoulhon, E., Grandclement, P., Taniguchi, K. et al., 2001,
Phys. Rev. D 63, 064029 (gr-qc/0007028)

\bibitem{}Heger, A., Woosley, S.E., Langer, N., Spruit, H.C., 2003,
astro-ph/0301374

\bibitem{}Houser, J.L, Centrella, J.M., Smith, S.C., 1994,
Phys. Rev. Lett. 72, 1314

\bibitem{}Iben,  I., Tutukov, A.V., 1996, ApJ 456, 738 

\bibitem{}Imshennik, V.S., 1992, PAZh 18, 489

\bibitem{}Imshennik, V.S., Nadezhin, D.K., 1992, PAZh 18, 79

\bibitem{}Imshennik, V.S., Popov, D.V., 1998, Astron. Lett. 24, 206 

\bibitem{}Imshennik, V.S., Ryazhskaya, O.G., 2004, Astron. Lett. 29,
831 (astro-ph/0402191)

\bibitem{}Kaplan, D.L., Frail, D.A., Gaensler, B.M., et al., 2004, 
astro-ph/0403313

\bibitem{}Langer, N., Yoon, S.-C., Petrovic, J., Heger, A., 2003,
astro-ph/0302232

\bibitem{}Mathews, G.J., Wilson, J.R., 1997, ApJ 482, 929 (astro-ph/9701142)

\bibitem{}Mathews, G.J., Wilson, J.R., 2000, 
Phys. Rev. D 61, 127304 (gr-qc/9911047) 

\bibitem{}Menou, K., Perna, R., Hernquist, L., 2001, ApJ 559, 1032
(astro-ph/0102478)

\bibitem{}Oechslin, R., Rosswog, S., Thielemann, F.-K.,  2002, Phys. Rev. D
65, 103005

\bibitem{}Olinto, A.V., 1987, Phys. Lett. B192, 71

\bibitem{}Pavlov, G.G. et al., 2002, ApJ 531, L53

\bibitem{}Podsiadlowski, Ph., Langer, N., Poelarends, A.J.T., et al. 2003, 
astro-ph/0309588

\bibitem{}Timmes, F. X., Woosley, S. E., Weaver, T. A., 1996, ApJ 457, 834

\bibitem{}Woosley, S.E., Heger, A., Weaver, T.A., 2002,
Reviews of Modern Physics 74, 1015

\bibitem{}Xu, R.X., 2004, astro-ph/0402659

\end{thebibliography}
\end{document}